# Decentralized Traffic Management
# Strategies for Sensor-Enabled Cars

Ziyuan Wang

Supervised by Dr. Lars Kulik and Prof. Kotagiri Ramamohanarao

National ICT Australia

Department of Computer Science and Software Engineering

The University of Melbourne, Australia

Email: ziyuanw@csse.unimelb.edu.au

29 June 2007

# Contents

# Chapter 1

# Introduction

Road traffic congestion is a major challenge nowadays. Building more roads is often not a viable solution, in particular in metropolitan areas where space is limited. The integration of sensing, communication and local computing within cars can be exploited to optimize transportation systems. Currently, a variety of automotive sensors are available to collect data related to a vehicle and its surroundings. In addition, communication techniques, especially, the *Dedicated Short Range Communications* (DSRC) support the information exchange between cars. In the near future, sensor-enabled cars will have a significant market share.

Considering the current trends in this domain, sensor-enabled cars can provide a new opportunity to improve road throughput, and thereby reduce traffic congestion. The congestion does not occur only because of the excessive traffic demand over the road capacity but also because of the inefficient utilization of roads and other factors such as traffic accidents, that decrease the traffic throughput. Considering the entire road network, some roads are underutilized while some are overloaded. There are two types of imbalances which effect the throughput negatively: (1) global factors impacting the entire network and (2) the local factors that are responsible for local perturbations. Local perturbations can result from an improper driving car whose impacts may get amplified along the road, and thus leading to a reduced traffic throughput. For example, if a car pushes into a small gap to change lane, the impact may appear small locally but is globally significant as the following cars may need to slow down considerably. This is known as "slinky-type effect" [4], which can lead to traffic congestions and car accidents.

Our work, therefore, focuses on optimizing traffic throughput when there are conflicting traffic flows, such as at intersections where a ramp leads onto the highway, lane changing when there are obstacles on the way. We set out to explore the benefit of applying simple traffic control rules combined with sensor-enabled cars to improve throughput in entire road networks. We attack the problem by addressing the following three research areas: (1) traffic merging algorithms at intersections and on-ramps; (2) robustness of algorithms not only at the intersections but also traffic jams result of some cars that are not sensor-enabled and drivers who do not obey the rules; and (3) traffic scheduling algorithms on the road network level.

## 1.1 Motivation

In the near future we will see considerable change in the driving task and the way traffic is managed. These changes are driven by the urgent need to address the serious problems of increased traffic congestion, energy waste, high fatality and injury rates, and environmental pollution.

We propose collision-free strategies for efficient traffic control, which lead to a significant improvement of the traffic throughput. As an example of motivating the proactive traffic control strategy, we consider the following highway merging situation. A positive side-effect of our strategy is a small number of local perturbations, i.e., a smaller number of speed changes, which reduces fuel consumption and air pollution. Using sensed information, the merging process will be "smoother" because a car on the ramp can adjust its speed earlier to adapt to the gaps between the cars on the main road and their speeds. This leads to smaller interruptions for the main road flow. Figure 1.1 shows a scenario where ramp cars merge onto a main road. In both cases we assume the same initial configuration (middle figure). In the upper figure we assume a priority-based merging algorithm where a car does not adapt its speed before it arrives at the merging section. This requires car x to slow down considerably in order to merge onto the main road. In the lower figure, we assume that car x adapts its speed before its arrival at the merging section and can merge immediately when it arrives at this section. This leads to a smaller impact on both traffic streams as the merging car has the same speed as the cars on the main road when it merges.

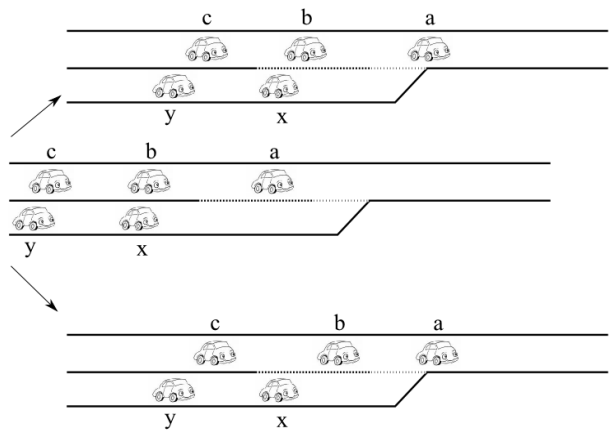

Figure 1.1: Example of proactive traffic control for sensor-enabled cars

The flowing challenges for traditional traffic management strategies motivate us to take an alternative approach:

- The size of the problem for the whole traffic network is very large.

- Measurements of traffic conditions are mostly local (via inductive loop detectors).

- Many unpredictable and hardly measurable disturbances (e.g., incidents, illegal parking, pedestrian crossings, and intersection blocking) may perturb the traffic flow.

The first challenge is the main reason that we adopt decentralized system design which is inherently scalable. The second and the third challenges can take advantage of sensor-enabled cars that facilitate sensing and communicating an extensive set of information.

## 1.2   Research Problem

The main objective of our research is to optimize the throughput of different traffic streams using sensor-enabled cars.

Decisions made by individual drivers concern when to accelerate or brake, to overtake or to enter a busy multi-lane road, under the constraints imposed by physical limitations and traffic rules. Our research assist drivers in making more informed and efficient decisions. The algorithm is based on the fact that sensor-enabled cars have more spatial information, such as location, speed, and acceleration or deceleration. On the other hand, to know better increases the cost in terms of consuming more sensors, computation power, and communication bandwidth. Therefore, we investigate: (1) what is the minimal set of spatial parameters that assess the traffic situation and how to identify a set of rules optimizing traffic locally; (2) how to deal with errors of sensor measurement and human factors; (3) how early should the algorithm make decision; (4) how will local decisions impact global performance.

We approach the research problem in three phases:

**Phase 1** We develop algorithms that proactively merge different traffic streams aiming at optimizing the traffic throughput. We evaluate what benefits we get from using the sensor-enabled cars.

**Phase 2** We investigate the efficacy of the algorithm in terms of robustness considering the human factors (driver reaction times, rule-breaking behaviors, etc.), the current penetration of sensor-enabled cars in the vehicle market, and the technical issues such as the accuracy of sensor measurement.

**Phase 3** We investigate the efficiency of traffic control strategy on a global level, in particular for a city road network that consists of extensive number of intersections and highway network, which consists of multiple on-ramps and exit-ramps as well as intersections with arterial roads. We evaluate the impact of the local optimal algorithm based on the study of first two phases.

## 1.3 Contributions

We have made the following contributions so far:

- We have proposed proactive traffic control algorithms that aim to use the current road facilities efficiently, including the improvement of overall traffic throughput, decrease of the travel time and economic fuel consumption. Conventional road traffic control strategies, such as intersection signal control, ramp metering, and variable message signs will benefit from our work.

- We have evaluated a range of algorithms using the following performance criteria for traffic control: latency, throughput, and fuel efficiency. We investigate what criteria can better evaluate the performance of traffic control algorithms.

- We have designed a controlled simulation environment intended to test various traffic control strategies. Using this, extensive empirical studies of the behavior of proposed algorithms can be provided.

We aim to achieve the following contributions in the future:

- We will refine the traffic control algorithms, which are more adaptive to a variety of traffic conditions.

- We will address the issue of robustness, which is a major challenge for the traffic control algorithms.

- We will also attempt to conduct more realistic simulations, such as a diversity of vehicle types who have different dynamics, i.e., speed, acceleration and deceleration abilities; different traffic patterns, such as Poisson arrivals and highly bursty traffic.

## 1.4   Structure

The rest of this report is structured as follows. In Chapter 2, we provide the background and related work on traffic flow control, with emphasis on merging algorithms. In Chapter 3, we describe the preliminary work: three proactive strategies for merging different traffic flows at intersections of the main road and the ramp. We present the simulation results of benefits we get from proactive merging algorithms and the shortcomings of these algorithms. Chapter 4 discusses possible future research directions and the expected timeline.

# Chapter 2

# Literature Review

Our approach to improve the traffic throughput relies on sensor technologies. The assumptions we make, the parameters we set for experiments are based on the current development of automotive sensors and communications. Therefore, we first give an overview on automotive sensors in Section 2.1. In Section 2.2, we present taxonomy of different approaches for traffic control. We show two different types of control design, centralized and decentralized systems. We go for the decentralized design by highlighting the strength and weakness of each design. In Section 2.3, we review scheduling algorithms in operating systems and packet switching networks considering their similarities with the problem of road traffic control. Since traffic flow has been intensively studied by means of mathematical models and simulation, in Section 2.4 we highlight current traffic models and the *intelligent driver model* (IDM), which is the foundation of our simulator.

## 2.1   Automotive sensors and DSRC

Recent advances in micro-electro-mechanical systems (MEMS) technology and wireless communications have enabled the development of low-cost, multifunctional sensor nodes that are small in size and with short-range wireless communication capabilities [1]. These tiny sensor nodes, which comprise sensing, data processing, and communicating compo-nents, can be installed to modern cars to facilitate traffic control. This new type of sensor-enabled cars are able to sense information about its own location and local traffic conditions, process this information, and communicate this information to other vehicles in its neighborhood. Fleming [8] gives a comprehensive up-to-date survey of automotive sensors. Li *et al.* [19] summarize the current research on intelligent vehicles. The current development of wireless communication techniques, especially, the *Dedicated Short Range Communications* (DSRC) designed for vehicular networks have a small communication range (basically, four to five hundred meters), thus we assume that sensor-enabled cars can communicate to five to eight cars in a highway scenario. Automotive sensors are typically required to have the total error less than 3% over their entire measured range, according to [8], thus we set the error range to this value when we test the robustness of the algorithms.

## 2.2   Traffic Control

Researchers have applied different techniques to improve traffic flow control, especially in the context of merging on-ramps and main roads traffic. These include queueing analysis, ramp metering, and automation control. Cowan [6] addresses the traffic merging problem from a queueing theory perspective and points out that arbitrary merging strategies result in the same average delay under equilibrium conditions. However, this is not sufficient for effective traffic control strategies. The merging section is not always under equilibrium conditions, so different merging strategies influence on the traffic throughput. The ramp metering strategy [21] assumes fully stop one stream. We can achieve safe merge without a full stop by arranging proper position and velocity of cars from two streams. Varaiya [26] argues that only full automation can achieve significant increase of traffic capacity. However, the shortcoming is that it needs major changes to the existing highway infrastructure.

Previous works on merging control algorithms rely on automatic control by calculating a proper speed and time to get to the merging point [20, 16, 27]. Adaptive cruise control (ACC) is partial automation control, which provides automation only in the longitudinal direction. Kesting *et al.* [17] state that ACC equipped cars can alleviate traffic congestion by simulating ramp cars merging onto the main road. Their approach does not address the impact of different merging strategies. Davis [7] proposes cooperative merging strategy to increase throughput and reduce travel times. In his strategy, a ACC equipped car adjusts its position according to the front car on the other road to create a safety gap without slowing down sharply. However, it only considers the nearest front car and thus the benefit is limited.

In theory, traffic control systems can be categorized into centralized and decentralized systems. In practice, there are no pure centralized systems. The present research mainly adopts a hybrid system architecture, i.e., a combination of a centralized and decentralized approach [26, 2]. Varaiya [26] discusses pure centralized and decentralized designs: a centralized has a tremendous computation and communication cost whereas a decentralized approach requires intelligence, which is expensive. In his approach, cars are grouped into tightly spaced train-like platoons and the first car of the platoon is controlled centrally whereas within platoons cars are decentralized. We will investigate whether inexpensive sensors in combination with simple traffic rules are sufficient for improving traffic flow. In our work, however, cars are not grouped into platoons to gain flexibility. Current research achieves some level of decentralization whereas we aim to build a fully decentralized system to optimize the traffic throughput.

Table 2.1: A comparison of traffic control strategies

|  | **Architecture** | **Group** | **Strategy** |
|---|---|---|---|
| **PATH** | H | P | Fully Automation |
| **VGrid** | H | P | ACC |
| **We** | D | I | Advice |

Table 2.1 summarizes the traffic control strategies we discussed from the aspects of system architecture (H denotes hybrid, D denotes decentralized), group or individual (P denotes platoon, I denotes individual) and control strategy.

## 2.3 Scheduling Algorithms

Since cars share a common resource: space, they are in competition for that resource. The problem of optimizing different traffic streams at intersections or merging scenarios can be mapped to the scheduling of jobs that are competing for mutually exclusive resources in the data packet scheduling in computer networks and job scheduling in an operating system. This analogy opens the door to a rich source of potential algorithms and analysis techniques that can be applied to the road traffic control strategy.

There are numerous algorithms for making the scheduling decision among all ready processes in operating systems. The simplest is first-come-first-served (FCFS), which selects the process that has been waiting the longest for service. Other scheduling algorithms include round-robin, which uses time-slicing to limit the running process to a short processor time and rotate among all ready processes; shortest process next, which selects the process with the shortest expected processing time; and shortest remaining time, which selects the process with the shortest expected remaining process time.

The commonly used criteria can be categorized to user-oriented and system-oriented. User-oriented criteria include turnaround time, response time, deadlines, and predictability. System-oriented criteria include throughput, resource utilization, fairness, enforcing priorities, and balancing resources. These criteria cannot be optimized all at the same time. For example, providing good response time may require a scheduling algorithm switching processes frequently. This increases the overhead of the system, thus reducing throughput. Therefore, the design of a scheduling algorithm is to find a compromise among competing requirements.

Irani *et al.* [14] model the conflicts between jobs with a conflict graph for developing traffic signal control at intersections. In a conflict graph, each node represents a type of job. If two types of jobs demand a common resource, there is an edge between those nodes in the graph. If there are two job of the same type in the system, one must wait until the other is completed. Thus, the set of jobs currently being executed must belong to nodes which form an independent set in the graph. The turnaround time of any job is bounded. Shah *et al.* [22] use real-time scheduling techniques to coordinate the movement of vehicles along intersecting roads.

The major difference between our work and the related work is that the "job" and "server" in our model are better informed by using sensor-enabled cars. We will investigate the benefit of more information in making decisions.

## 2.4   Traffic Model

Traffic models and simulations have provided insights into understanding traffic phenomena in order to eventually make decisions which may alleviate congestion and optimize traffic flow. Various tools and techniques such as kinetic gas theory, fluid dynamics, and cellular automata have been applied to model traffic (for a comprehensive review, see [12]). Traffic models and simulations can be categorized into three classes: microscopic (particle-based), mesoscopic (gas-kinetic), and macroscopic (fluid-dynamic) models according to the level of detail of the simulation. Macroscopic models are primarily concerned with traffic situations involving a large number of vehicles, interested in the collective behavior of average traffic variables (i.e., density and flow). In microscopic models, the movement of individual vehicle is simulated. In mesoscopic models, aspects of both macroscopic models and microscopic models are combined. Microscopic models are generally thought to be more suitable for evaluating the relation between traffic flow and behavior of the individual vehicle. We also adopt this microscopic approach in our research.

The *intelligent driver model* (IDM) [24] is a microscopic traffic model. Vehicles tend to approach the maximum velocity and maintain safety distance to the front vehicle. The safety distance depends on the following car's velocity and velocity difference from the front car. The acceleration and deceleration depend on its own velocity, safety distance and actual distance to the front car. We build a simulation environment based on this model because it has the following advantages [23]: the model is a realistic description of both the individual driving behavior and collective dynamics of the traffic flow; many aspects of traffic control strategies can be simulated by representing different driving styles, which are easy to implement.

# Chapter 3

# Proactive Merging Algorithms

In our first work on road traffic optimization by using sensor-enabled cars we address the issue of how to optimize traffic throughput on highways, in particular for intersections where a ramp leads onto the highway. In this chapter, we present proactive traffic control algorithms for merging different streams of sensor-enabled cars into a single stream. Sensor-enabled cars allow us to decide where and when a car merges before it arrives at the actual merging point. This leads to a significant throughput improvement for the traffic as the speed can be adjusted appropriately. An extensive set of experiments shows that proactive merging algorithms outperform the priority-based merging algorithm in terms of throughput and latency.

## 3.1 Introduction

The merging section of highways is a bottleneck that influences the traffic throughput significantly. There is already a considerable amount of research, in particular approaches from queuing theory or statistics [6]. However, this research did not consider sensor-enabled cars nor compared different merging strategies under realistic simulation environments. Our work, therefore, focuses on optimizing traffic throughput when merging different traffic flows at intersections by using sensor-enabled cars. We set out to explore the benefit of applying simple traffic control rules at the merging section if all cars are sensor-enabled.

The key insight is that dissociating the decision point and the actual merging point can optimize traffic throughput. Currently, most of the literature [3, 16, 20] assumes that the decision point and the actual merging point coincide based on the fact that traditionally, a driver arrives at the merging point and makes a decision of how to merge at that point. The reason that we are able to make proactive decision is that sensor-enabled cars can obtain necessary information for safe merging much earlier than normal cars.

To compare our merging strategies, we outline a priority-based, non-proactive merging algorithm who serves as the benchmark algorithm. This is a strategy commonly adopted in the current merging intersections. Priority-based means the cars from one stream always have the right of way, e.g., a car on the ramp must give way to the cars on the main road. The ramp car only merges when the safety gap on the main road is enough. Meanwhile, the cars on the main road ignore the ramp car and do not create gaps for the ramp car to merge. Non-proactive means a car on the ramp does not adjust its speed before it arrives at the actual merging point.

## 3.2   Algorithm Overview

The basic idea behind the algorithm is to use the knowledge of position, velocity, and acceleration received beforehand in making merging decisions. Each car knows about a small number of cars in the neighborhood. When a car arrives at the decision point, which is before the actual merging point, it chooses a proper gap to prepare for merging. Along the path of approaching to the merging point, it adjusts velocity to catch that gap when it arrives at the actual merging point. In such way, the velocity change is small compared with the non-proactive strategies.

A car on the ramp approaches the merging point as if there is a car stopping at the end of the merging section. Therefore, its velocity is composed of a decreasing tendency and the adjustment to the gap on the other stream. The requirement for the merging algorithm is that a safety distance should be guaranteed not only along each road before merging but also at the point of and after the merging maneuver. The objective of proactive merging is to make decision early so that a ramp car can merge to the main road without slowing down significantly.

The proactive merging algorithm works as follows:

**phase 1** make decision of which car merges first at the decision point considering the traffic condition rather than which road the car is on, in contrast to priority-based algorithm.

**phase 2** adjust the velocity prior to the merging point. This is the key feature of proactive merging algorithm.

Several different algorithms can be used to decide the merging order. Examples of such algorithms are distance-based, velocity-based, load-based, increase-based, or some combination of these. In Table 3.1, we summarize distance-based and velocity-based merging algorithms, which we implement and compare against priority-based merging algorithm.

Table 3.1: Traffic Merging algorithms overview

|  | Knowledge | Right of Way | Assumption |
|---|---|---|---|
| **Distance-based** | Position | the car that is closest to the merging point | velocity does not vary much |
| **Velocity-based** | Position, velocity | the car that arrives to the merging point first | acceleration does not vary much |

Each algorithm performs better than others under certain traffic condition. After we examine the performance of different algorithms we develop a more refined, adaptive algorithm. It switches to the strategy that is best fit for the current traffic situation. This approach incrementally improves on the throughput that specifically suits the unstable traffic scenario.

Furthermore, the concept of *sliding decision point* will make the proactive merging algorithm more adaptive. The position of decision point can be changed according to different road conditions. If the traffic condition changes frequently then it is better to make the decision point closer to the actual merging point because it is difficult to predict the changes.

### 3.2.1 Pseudocode

In Figure 3.1, we explain the notation we use in the following text and pseudocode (Algorithm **??**). The main road and the ramp meet in the merging area. We denote the start point by O and the end point by E, which coincides with the end of the ramp. The decision point is denoted by D, which indicates the point at which the car decides where and when to merge between O and E. The decision point D is not fixed and can be adjusted according to the traffic. In our experiment, we analyze how the velocities of the cars and the distance between the decision point and the merging point affect the traffic flow. A car can adjust its decision point in the gray area to optimize the traffic flow. We test whether or not D is the function of velocity in the following sense: if the velocity is high, D should be far away from O; if the velocity is low D should be near to O. We expect that the distance between D and O (denoted by $\overline{DO}$) is a key parameter that impacts the performance of merging algorithms. If $\overline{DO}$ is zero, it means that not making decision early implies in no benefit from separating D and O. As $\overline{DO}$ increases, the benefit will increase but after a point the benefit increase becomes trivial because making a decision too early is not adaptive to the varying traffic conditions.

A car is characterised by the following attributes: position, velocity, and acceleration. A car list is a logical concept for a number of cars that share common characters. For example, the three car lists: *RampList*, *MainList*, and *OutList*, which means the ramp cars that have not arrived at the merging point, the main road cars that have not arrived at the merging point, and the sequence of those two groups of cars after they pass the merging point computed by merging algorithms. Based on the definition, $RampList$ is $\{x, y\}$, and $MainList$ is $\{c, d, e\}$ in Figure 3.1. $OutList$ varies depending on the merging algorithm. It may be $\{c, d, x, e, y\}$ in priority-based merging algorithm, $\{c, x, d, y, e\}$ in distance-based merging algorithm, $\{x, c, d, y, e\}$ in velocity-based algorithm if car x with much higher velocity than car c. We show how the velocity-based merging algorithm work in Algorithm **??**.

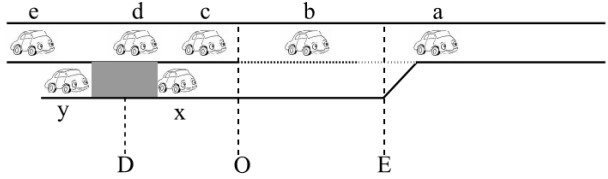

Figure 3.1: Denotation of merging

## 3.3   Performance Evaluation

In this section we compare the performance of distance-based, velocity-based, and velocity-proactive merging algorithms (referred to as D, V, and PV in the figures) against priority-based algorithm (referred to as R) in a variety of simulation settings.

### 3.3.1   Performance Metrics

In the literature [7, 2, 5], delay, traffic flow and capacity are usually taken as key criteria of traffic control strategies. We use these metrics for evaluating the traffic merging algorithms discussed in this paper. Note that these are dependent on each other, and it is impossible to optimize all of them simultaneously.

**Latency** This is the time to fill up a certain number of cars. It describes how quickly the system absorbs incoming cars. Our aim is to minimize for both streams the waiting time over a period.

**Throughput** This is the number of cars that complete merging over a period of time.

**Flow** This is defined by the product of density and velocity. The maximum of traffic flow occurs at some density (with a corresponding velocity). The maximum traffic flow is called the capacity of the road.

**Acceleration and deceleration** This is a main factor to impact fuel consumption [4]. In addition, it affects the passengers comfort. That is why sharp acceleration and braking had better be avoided.

### 3.3.2 Simulation

We developed a Java-based simulation environment based on an existing microscopic traffic simulation model called *intelligent driver model* (IDM) [24]. The IDM parameters for the simulations are given in Table 3.2. These values are known to reflex realistic traffic conditions. We settle for a relatively high velocity (100 km/h) because the higher velocity the more quickly to distinguish a good merging algorithm from a bad one. The set of acceleration and deceleration are lower than the physically possible to avoid collisions, moreover, to make the algorithms restrictive for the same purpose of setting high velocity value: to distinguish strategies more quickly because the smaller of the value the easier for a strategy to break. However, the benefit from small values of acceleration and deceleration is the decrease of fuel consumption. Safety distance varies with the velocity difference of two following cars. Minimum distance is the distance between two standstill cars.

All the merging strategies use the same simulator configuration but differ in terms of (a) the type of information they use in making merging decisions, and (b) the position at which they make the decision. We compare the algorithms with respect to latency, throughput, and average velocity.

Table 3.2: IDM parameters

| Parameter | Value |
|---|---|
| Desired velocity | $100\ km/h$ |
| Safe time headway | $1.5\ s$ |
| Maximum acceleration | $1\ m/s^2$ |
| Maximum deceleration | $3\ m/s^2$ |
| Minimum distance | $2\ m$ |

We evaluate the performance in the closed system without cars going out to investigate what are the key parameters from the set that includes various spacial parameters. We conducted a pre-study to identify those initial settings that can quickly distinguish the performance of merging algorithms. We narrow down to four parameters: initial density of the main road, incoming rate of ramp cars, the decision point, and the ramp length. We vary only one parameter in each simulation run and keep the others constant. The initial settings are combinations of the light, medium, and heavy traffic on the main road and on the ramp (see Table 3.3). We set a maximal incoming rate of ramp cars as 12 cars per minute because the capability of a ramp is limited. The incoming rate is set to be constant for simplicity at first and then set it following a Poisson distribution to better reflect real traffic situation. The length of main road is 10 kilometers; the ramp is 400 meters in all experiments except when we test the impact of ramp length, so we shorten it to 200 meters.

Table 3.3: Experiment settings

| | Light | Medium | Heavy | unit |
|---|---|---|---|---|
| **Main road** | 5 | 10 | 15 | $cars/km$ |
| **Ramp** | 6 | - | 12 | $cars/minute$ |

## 3.4    Results

In general, advantages of proactive merging algorithms over the priority-based algorithm are the following: First, the latency in proactive merging algorithms decreases by third compared with the priority-based algorithm when the loop is saturated. Cars on the ramp can merge into the main road more quickly. Second, the traffic flow in proactive merging algorithms is higher before the main road is saturated in the priority-based algorithm. As long as the average velocity is above 20 m/s (72km/h), the traffic flow is higher (Figure 3.2, Figure 3.3, and Figure 3.4). This is because traffic flow is defined by the product of the average velocity and the density; there are more cars on the main road in proactive merging algorithms due to less latency than the priority-based algorithm. At first, the larger number of cars overcomes the slightly smaller average velocity, as more cars in, the main road starts getting saturated, the average velocity drops to satisfy smaller distance. In addition, Figure 3.5 shows that the impact on acceleration is smaller for proactive merging algorithms in most traffic conditions.

The disadvantage is the average velocity in proactive merging algorithms is slightly lower. This is the side-effect of larger number of cars merging into the main road. In consequence, the distance between cars gets smaller, which leads to lower velocity because velocity is in proportion of distance due to safety requirement. Nevertheless, this lower velocity is the cost of higher throughput achieved.

We must manage complex tradeoffs among factors, such as velocity, throughput, and latency. The traffic flow increases at the beginning as the ramp cars merge, and decreases as the average velocity decreases. In order to maintain a relatively high traffic flow as well as a large number of cars get into the loop, we need to find a balance between the velocity and the number of cars merge into the loop.

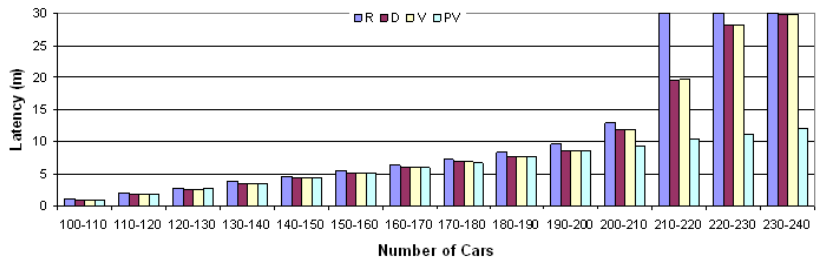

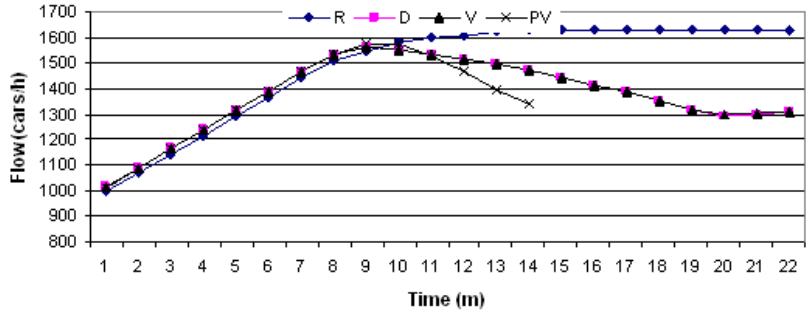

Figure 3.2: Medium initial density of the main road

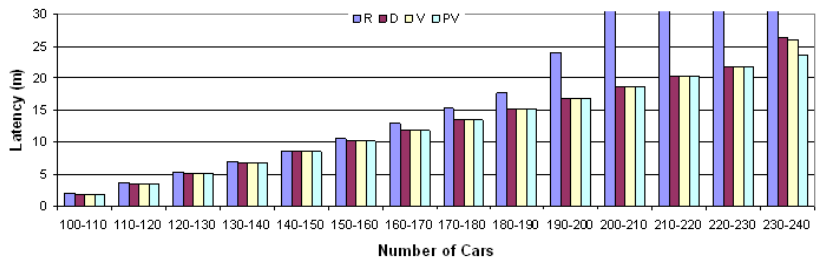

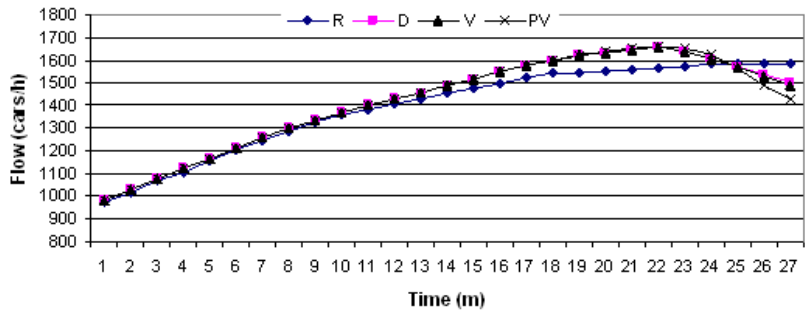

Figure 3.3: Medium initial density of the main road with low incoming rate of the ramp

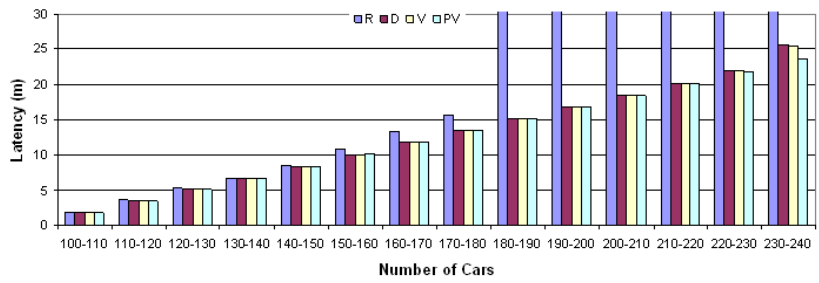

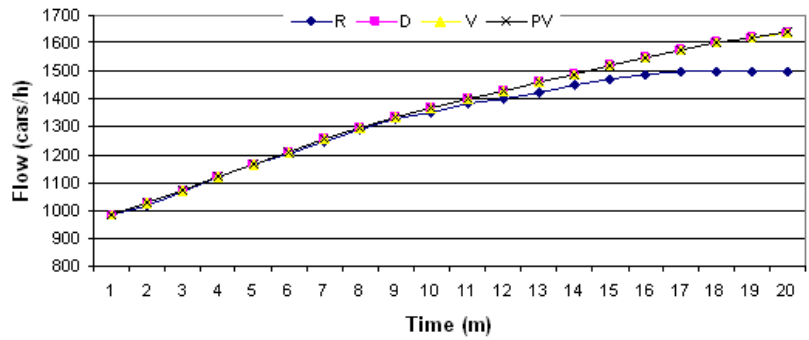

Figure 3.4: The performance of merging algorithms with a short ramp

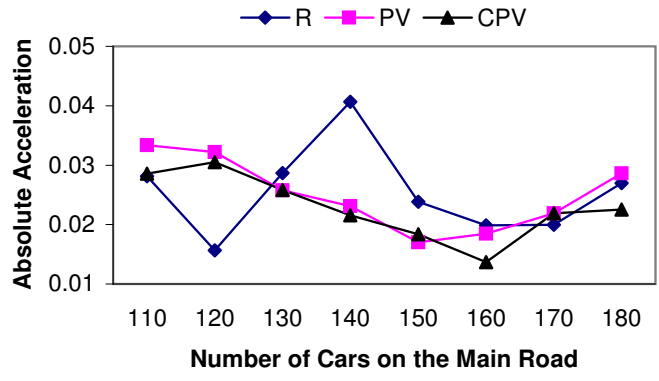

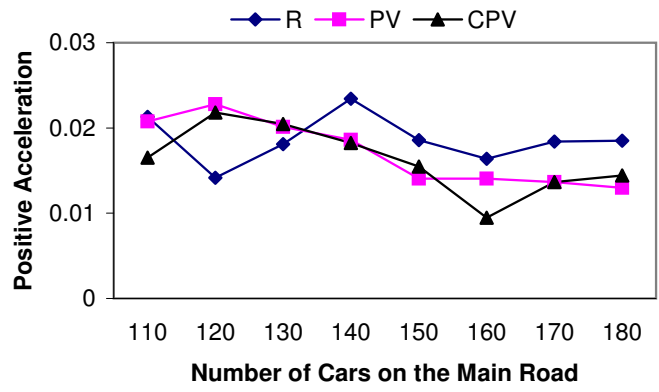

Figure 3.5: The impact on acceleration

# Chapter 4

# Future Directions

## 4.1 Research Questions

There are several research directions that we plan to follow:

### 4.1.1 Higher degree of realism

Considering a comprehensive set of traffic conditions, such as obstacles in the way, highly bursty traffic (e.g, burst of traffic in rush hours), multi-lane, and different types of vehicles, we will implement the following in our simulations to achieve a higher level of realism:

**Obstacles** When there is an obstacle in the way, the traffic stream heading to the obstacle need to change lane. This lane-changing action involves coordinate two conflicting streams of cars. It is different from the highway and ramp merging scenario because in this case there is no actual merging point, which means the car of the blocked stream can merge at any position before it arrives at the obstacle. The question is to when to merge can optimize the throughput of the two streams.

**Traffic pattern** We previously assume that the traffic density is uniform initially and the rate of incoming cars to the system is constant. However, there are highly bursty traffic situations, such as many cars arriving or leaving a certain place at the same time (e.g, a stadium, a university). To reflect the traffic condition more realistic, we will set the rate of incoming cars to follow the Poisson process.

**Multi-lane** We will extend our simulations from single lane to multiple lanes.

**Heterogeneity** We will also extend from single type of vehicles to different types of vehicles.

## 4.1.2 Robustness of algorithms

We will address these following issues regarding the robustness of the traffic control strategies:

**Imperfect information** In previous work, we assume that the spatial information which sensor-enabled cars get is accurate and precise. Automotive sensors are typically required to have the total error less than 3% over their entire measured range, according to [8]. Hernandez [13] also reports that when the car speed is lower than $5km/h$ sensors give very corrupted and misleading information.

**Human factors** A traffic control algorithm also needs accommodate various human factors, such as reaction times, and be prepared to handle drivers who do not obey the control rules.

Treiber *et al.* [25] include essential aspects of driver behavior in their traffic models, specifically, estimation errors. They model estimation errors for the net distance and the velocity difference to the preceding vehicle. The estimation errors are modeled as stochastic Wiener processes [9] and lead to time-correlated fluctuations of the acceleration. This model is useful when we consider the cars without certain type of sensor.

### 4.1.3 Development of theoretical model

In our research, a microscopic car-following model (see Section 2.4) is used as the underlying mobility model. Since mobility modeling in vehicle control field is a very sensitive issue, the different modelings with different level of granularity might generate the different results and different conclusion. A detailed discussion to justify why such a microscopic mobility modeling can serve the purpose of faithfully validating the proposed algorithm.

We will provide theoretical support for the simulations. For example, we may develop a theoretical model in which a stream of vehicles with random spacing could be made more regular by the proposed strategy and this regularity could be analytically derived.

A greater generalization of the traffic merging algorithm will be applied in general traffic control. In particular, we are interested in studying the performance of different strategies in road crossings without traffic lights, such as roundabout. Since sensor-enabled cars can get more information and can communicate among each other, we plan to investigate the performance of roundabout road intersections.

Kakooza *et al.* [15] propose a mathematical model to analyze the different types of road intersections. Their simulation results indicate that under light traffic, roundabout intersections perform better than signalized and unsignalized in terms of easing congestion; under heavy traffic, signalized intersection perform better in terms of easing traffic congestion compared to unsignalized and roundabout intersections.

### 4.1.4 Study in local decisions' impact on global performance

How to optimize traffic flow at network level that involves a number of intersections at urban road networks and multiple on-ramps and exit-ramps at freeway networks. We will investigate how to exploit scheduling algorithms in the context of traffic flow management at the road network level.

The algorithm described in Chapter 3 is based on the local negotiation of two vehicles

(one on the main road, one on the freeway entrance road). It is well known that sometimes the algorithm achieving the local optimization may end up to the global sub-optimization or result in the system oscillation. Whether the proposed proactive merging algorithm will suffer from this is not clear at this moment.

Researchers have agreed on that the traffic control problem on a global network level is practically unsolvable by traditional optimization techniques (see, e.g., [18, 11]) because of exponential complexity of the algorithms. Hence, a number of decentralized optimal strategies whose actions are coordinated heuristically by a superior control layer have been proposed [10]. Lammer *et al.* [18] present a self-organizing, decentralized control method for global coordination of traffic signal control. They map the problem to phase-oscillator models. By synchronizing these oscillators, the desired global coordination is achieved. The concept applies to networks where time sharing mechanisms between conflicting flows in nodes are required and where a coordination of these local switches on a network level can improve the performance. Gershenson [11] proposes self-organizing methods using simple rules to coordinate traffic lights to improve traffic flow.

The key research problem underlying all the above questions is to search for minimal sets of rules which optimize traffic throughput.